\documentclass{pasj00}

\begin{document}
\SetRunningHead{N.~Kikuchi, T.~Nakamoto, and K.~Ogochi}
{Disk--Halo Model for Flat-Spectrum T Tauri Stars}
\Received{1999 April 23}
\Accepted{2002 June 16}

\title{Disk--Halo Model for Flat-Spectrum T Tauri Stars}

\author{Nobuhiro \textsc{Kikuchi}}
\affil{Earth Observation Research Center, National Space Development
  Agency of Japan, \\ 1--8--10 Harumi, Chuo-ku, Tokyo 104--6023}
\email{kikuchi@eorc.nasda.go.jp}

\author{Taishi \textsc{Nakamoto}}
\affil{Center for Computational Physics, University of Tsukuba,
  Tsukuba, Ibaraki 305--8577}
\and
\author{Koji \textsc{Ogochi}}
\affil{Information Technology of Japan Inc.,
  Mito, Ibaraki 310--0803}

\KeyWords{planetary systems: protoplanetary disk
--- radiative transfer
--- stars: formation}

\maketitle

\begin{abstract}
We explore the origin of the flat spectrum seen in some T~Tauri
stars by considering a three-component structure: a central star,
a circumstellar disk, and a dusty halo.
The radiative energy transport is faithfully treated by solving the
angle- and frequency-dependent radiative transfer equation in two space
dimensions assuming axisymmetry, and hence the radiative equilibrium
temperature in the disk and halo is determined simultaneously.
The disk is effectively heated by the scattering and reprocessing of
stellar radiation through the halo.
The large mid- to far-infrared excess originates from the photosphere of
the warmed disk, resulting in a flat spectrum, as observed.
The halo which we consider is observed as a compact reflection nebula,
and is discriminated from extended, disk-like envelopes around
flat-spectrum T~Tauri stars.
We show that the overall spectral shape of flat-spectrum T~Tauri stars
can be reproduced by the present {\it disk--halo} model.
\end{abstract}

\section{Introduction}

It seems reasonable to suppose that the infrared excesses of T~Tauri
stars can be attributed to thermal dust emission from circumstellar
disks.
By modeling the observed spectral energy distributions (SEDs), one can
derive the disk properties, such as masses, radii, and temperature
distributions (Adams et al.\ 1988; Strom et al.\ 1989; Beckwith et al.\
1990).
Such spectral modeling, however, has revealed an extreme class of
T~Tauri stars, namely flat-spectrum T~Tauri stars.
The large mid- to far-infrared excesses of flat-spectrum T~Tauri stars
require their disks to have a temperature distribution of the form
$T \propto R^{-1/2}$ (Adams et al.\ 1988), where $R$ is the distance
from the rotation axis, whereas a $T \propto R^{-3/4}$ dependence is
predicted by both a standard accretion disk (Lynden-Bell, Pringle 1974)
and a spatially flat reprocessing disk (Adams, Shu 1986).
This implies that the disks of flat-spectrum T~Tauri stars are warmer in
the outer regions than predicted by the simple disk models; we must
therefore consider the mechanisms which heat the outer region of the
disk.
Disk flaring allows a disk to receive more emission from the central
star, and to produce temperature distributions shallower than $T \propto
R^{-3/4}$ (Kusaka et al.\ 1970; Kenyon, Hartmann 1987), but does not
suffice to reproduce the observed flux.

The infrared excesses may originate from another circumstellar dust
component.
Calvet et al.\ (1994) invoked infalling envelopes, which were originally
applied to the spectral modeling of protostars (Adams, Shu 1986;
Kenyon et al. 1993), and showed that the infalling envelopes can produce
the mid- to far-infrared excesses of flat-spectrum T~Tauri stars.
In fact, observations have shown that there exists an extended,
disk-like structure of radius $\sim 1000\ {\rm AU}$ around a typical
flat-spectrum T~Tauri star, HL~Tau (Sargent, Beckwith 1991).
Furthermore, Hayashi, Ohashi, and Miyama (1993) have revealed an
infalling motion in the disk-like structure of HL~Tau, suggesting that
it is a remnant of an infalling envelope.

Although the infalling envelope model of Calvet et al.\ (1994) is
successful in reproducing the flat spectrum, it does not take into
account the following two important effects.
First, as pointed out by Natta (1993), envelopes scatter and reprocess
the stellar radiation toward the disk, and thereby alter the temperature
distribution in the disk (see also Butner et al.\ 1994; D'Alessio et
al.\ 1997).
Second, the disk, itself, significantly influences the temperature
distribution in the envelope.
Since these two effects are coupled with each other, they cannot be
treated separately.
Hence, to elucidate the substantial mechanism for the flat spectrum, the
temperature structure of the disk and envelope should be solved
simultaneously.
In the present analysis, we consistently handle the radiative energy
transport in the disk and envelope by solving the angle- and
frequency-dependent radiative transfer equation in two space dimensions
by assuming axisymmetry.

In this paper, we demonstrate that the mid- to far-infrared excesses of
flat-spectrum T~Tauri stars can originate from the {\it disk}.
Since the envelope which heats the disk can be as compact as the disk
itself, we henceforth call it a {\it halo} to distinguish it from an
extended infalling envelope.
The reflection nebula of HL~Tau revealed by the Hubble Space Telescope
(Stapelfeldt et al.\ 1995) may be regarded as an observational
counterpart of the halo.

\section{Model}

The disk--halo model consists of a central star, a circumstellar disk,
and a halo surrounding both of them.
We first assume the axial and equatorial symmetry.
The central star has luminosity $L_\ast$ and mass $M_\ast$, and radiates 
as a blackbody with effective temperature $T_\ast$.
We consider models with $M_\ast = 0.5\ \MO$ and $T_\ast = 4000\ {\rm K}$,
which are typical values for pre-main-sequence stars in the
Taurus--Auriga molecular cloud.
We adopt $L_\ast = 5\ \LO$ for models without disk accretion, in which
all the energy is radiated by the central star.
Models with disk accretion are also examined, for which stellar
luminosity is taken to be $L_\ast = 2\ \LO$ and the disk, itself, has
some intrinsic luminosity.

The surface density distribution of the disk is assumed to be a power
law,
\begin{equation}
\sigma( R ) = \sigma_1 (R/{\rm AU})^{-3/2}~,
\label{eqn:sigma}
\end{equation}
in the range $1\ {\rm AU} \le R \le 100\ {\rm AU}$.
Outside this range, we make the surface density go to zero smoothly
inward down to $R_{\rm in} = 0.1\ {\rm AU}$ and outward up to $R_{\rm
out} = 120\ {\rm AU}$.
The value of $\sigma_1$ specifies the disk mass.
We consider models with $\sigma_1 = 2 \times 10^3\ {\rm g~cm}^{-2}$,
which gives $0.02\ \MO$ for the disk mass, and is consistent with the
minimum mass solar nebula (Hayashi et al.\ 1985).
For the given surface density distribution, the two-dimensional density
distribution in the disk is determined by the hydrostatic equilibrium in
the vertical direction under the gravity of the central star.

In addition to the radiation from the central star, viscous dissipation
also heats the disk if it is undergoing accretion, as observations of
classical T~Tauri stars suggest (e.g., Bertout et al.\ 1988).
We take into account the effects of disk accretion in a simple way as
follows.
Assuming that the mass accretion rate in the disk, $\dot{M}_{\rm d}$, is
constant with radius, we express the energy generation rate per unit
area as (Lynden-Bell, Pringle 1974)
\begin{equation}
D( R ) = \frac{3 G M_\ast \dot{M}_{\rm d}}{4 \pi R^3}
\left( 1 - \sqrt{ \frac{R_{\rm in}}{R} } \right)
\label{eqn:D}
\end{equation}
in the range $R_{\rm in} \le R \le R_{\rm out}$.
We further assume that the energy generation rate per unit mass does not
depend on the vertical height, which can be written as
\begin{equation}
\varepsilon( R ) = D( R )/\sigma( R )~.
\end{equation}
The intrinsic disk luminosity is then given by integrating equation
(\ref{eqn:D}) over the disk area as
\begin{eqnarray}
L_{\rm d}
& = & 2 \pi \int_{R_{\rm in}}^{R_{\rm out}} D( R ) R dR \nonumber \\
& = & \frac{G M_\ast \dot{M}_{\rm d}}{2 R_{\rm in}}
\left[ 1 - 3 \frac{R_{\rm in}}{R_{\rm out}}
+ 2 \left( \frac{R_{\rm in}}{R_{\rm out}} \right)^{3/2} \right]~.
\label{eqn:Ld}
\end{eqnarray}

In a classical picture of accretion disks, the disk extends down to
the stellar radius $R_\ast$ to form a boundary layer, where disk
material slows down from Keplerian to stellar rotation velocity.
Radiation from a boundary layer accounts for the ultraviolet and optical
excesses of T~Tauri stars (Lynden-Bell, Pringle 1974).
On the other hand, our disk model has an inner radius of $R_{\rm in} =
0.1\ {\rm AU}$, which is well above the stellar surface.
The underlying assumption for this is that the stellar magnetic field
truncates the disk at several stellar radii, and accretion from the disk
onto the central star takes place along the magnetic field lines
(Bertout et al.\ 1988).
A ``bright spot'' is then formed where the accreting material impacts
the stellar surface.
The luminosity of the radiation from this bright spot can be as large as
the stellar luminosity, and therefore it will also be important as an
additional energy source for disk heating.
In our model, disk material flows at the rate $\dot{M}_{\rm c}$ from the
disk inner radius onto the stellar surface, and the amount of radiation
energy emitted by the bright spot is given by
\begin{equation}
L_{\rm c}
= \frac{G M_\ast \dot{M}_{\rm c}}{R_\ast}
- \frac{1}{2} \frac{G M_\ast \dot{M}_{\rm c}}{R_{\rm in}}~,
\label{eqn:Lc}
\end{equation}
which derives from the assumption that the energy difference of
accreting material between the total kinetic + gravitational energy at
$R = R_{\rm in}$ and the gravitational energy at $R = R_\ast$ is
released by radiation from the bright spot.
The stellar radius, $R_\ast$, is defined by $L_\ast = 4 \pi R_\ast^2
\sigma_{\rm B} T_\ast^4$.
We do not assume that $\dot{M}_{\rm c}$ is equal to $\dot{M}_{\rm d}$,
i.e., we consider models with unsteady disk accretion.
For simplicity, the bright spot is assumed to radiate isotropically as a
blackbody with an effective temperature of 8000~K.

The density distribution in the halo is supposed to be a power law,
\begin{equation}
\rho_{\rm h}( r ) = \rho_1\ (r/{\rm AU})^{-p}~,
\end{equation}
for
$r \ge 1\ {\rm AU}$,
where $r$ is the distance from the central star and $\rho_1$ is a
reference density.
To avoid a singularity at $r = 0$, we take
$\rho_{\rm h}( r ) = \rho_1\ [1 + p/2 - p/2 (r/{\rm AU})^2]$
for
$r \le 1\ {\rm AU}$.
A different assumption for cutting off the power-law behavior, for
instance, keeping the power law down to 0.1~AU, did not change the
results significantly.
Since bipolar outflows may evacuate material along the rotation axis, we
introduce bipolar holes with an opening half-angle of $\theta_{\rm bp}$.
The halo density is reduced slowly and smoothly inside the bipolar holes
by multiplying a factor
$f_{\rm bp} = 3 (\theta/\theta_{\rm bp})^2 - 2 (\theta/\theta_{\rm bp})^3$
to the original halo density for $\theta \le \theta_{\rm bp}$, where
$\theta$ is the polar angle.

The frequency-dependent absorption and scattering coefficients of dust
particles are taken from Miyake and Nakagawa (1993), assuming that the
size distribution of dust particles, $n( a )$, with respect to the
radius, $a$, is expressed by
$n( a ) \propto a^{-3.5}$ for $0.01\ \mu{\rm m} \le a \le 1\ \mu{\rm m}$
and
$n( a ) \propto a^{-5.5}$ for $1\ \mu{\rm m} \le a \le 10\ \mu{\rm m}$.
Miyake and Nakagawa (1993) assumed that dust particles are composed of
silicate and water ice.
We further assume that the dust vaporization temperatures are 1000~K and
100~K for silicate and water ice, respectively.
For simplicity, we adopt the approximation of isotropic scattering.

Given the density distributions and the radiative and viscous heating
sources, we calculate radiative equilibrium temperature in the disk and
halo which satisfies the condition of radiative equilibrium,
\begin{equation}
4 \pi \int_0^\infty \kappa_\nu^{\rm abs} B_\nu( T ) d\nu
= c \int_0^\infty \kappa_\nu^{\rm abs} E_\nu d\nu + \varepsilon~,
\label{eqn:radeq}
\end{equation}
where $\nu$ is the frequency, $\kappa_\nu^{\rm abs}$ is the mass
absorption coefficient, $B_\nu$ is the Planck function, $T$ is the
temperature, and $E_\nu$ is the monochromatic radiation energy density.
To solve equation (\ref{eqn:radeq}), we adopt the variable Eddington
factor method in a cylindrical coordinate system $( R,Z )$ (Stone et
al.\ 1992), in which we integrate the time-dependent radiation moment
equations forward in time until a stationary state is achieved.
The radiation moment equations are closed by introducing variable
Eddington factors, which are in turn calculated from the solution of
the angle- and frequency-dependent radiative transfer equation.
For every timestep, the disk is assumed to be in hydrostatic
equilibrium determined by the temperature distribution at the previous
timestep.
Stationary state solutions obtained in this way then satisfy the
conditions of radiative and hydrostatic equilibria simultaneously.
We take the size of the computational box to be $200\ {\rm AU} \times
200\ {\rm AU}$ using a $100 \times 100$ non-uniform spatial grid, and
employ $203 \times 101$ angle rays and 101 frequency meshes.
The accuracy of our radiative transfer method has been confirmed by test
calculations described in Stone et al.\ (1992) and in Masunaga, Miyama,
and Inutsuka (1998).

\section{Results}

\subsection{Standard Case}

The halo density distribution,
$\rho_{\rm h}( r ) = \rho_1\ (r/{\rm AU})^{-3/2}$,
is a natural consequence of infalling material with a constant mass
infall rate, $\dot{M}$, if $\rho_1$ is given by
$\rho_1 = (4 \pi)^{-1} \dot{M} (2 G M_\ast)^{-1/2} ({\rm AU})^{-3/2}$.
From the spectral modeling of protostars in the Taurus--Auriga molecular
cloud, Kenyon et al.\ (1993) derived a typical mass infall rate of
$\dot{M} = 4 \times 10^{-6}\ \MO\ {\rm yr}^{-1}$, which gives
$\rho_1 = 3 \times 10^{-14}\ {\rm g~cm}^{-3}$
for 
$M_\ast = 0.5\ \MO$.
Noting this value of $\rho_1$, we first present results for a halo model
with $p = 3/2$, $\rho_1 = 10^{-14}\ {\rm g~cm}^{-3}$ and $\theta_{\rm
bp} = 60^\circ$.
In a standard case detailed in this subsection, disk accretion is not
included, and all of the energy arises from the central star with
luminosity $L_\ast = 5\ \LO$.

\begin{figure}
\begin{center}
\includegraphics[width=85mm,clip]{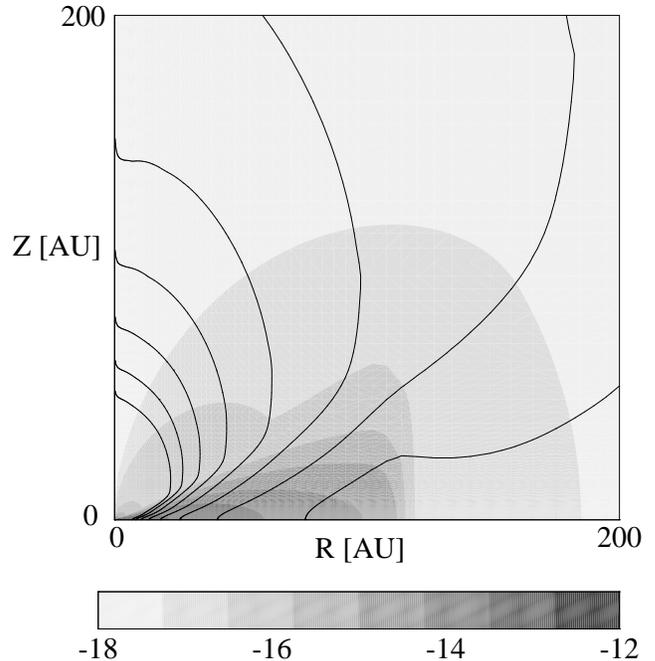}
\end{center}
\caption
{Density and temperature distributions for the standard model having
$\rho_1 = 1.0 \times 10^{-14}\ {\rm g~cm}^{-3}$,
$\theta_{\rm bp} = 60^\circ$ and $p = 3/2$.
The grey scale and contours represent the density and temperature,
respectively.
The innermost contour is $T = 100\ {\rm K}$, and the contours decrease
outward in $10\ {\rm K}$ intervals.}
\label{fig:fig1}
\end{figure}

Figure 1 shows the density and temperature distributions in the disk and
halo.
The temperature gradient is shallow along the axis of symmetry where the
density is very low, while it is steep along the midplane where the disk
is situated.
The resulting temperature distribution is far from spherically
symmetric, as illustrated by the contours in the teardrop shape.
It can also be seen from figure 1 that the disk is not isothermal in the
vertical direction, with a surface temperature higher than the midplane
temperature.
The vertical structure at $R = 10\ {\rm AU}$ is shown in figure 2a.
The disk surface lies at $Z \simeq 4\ {\rm AU}$, which is defined as the
position above which the density of the disk decreases below that of the
halo.
As illustrated by the solid line in figure 2a, the temperature $T$
begins to rise at $Z \simeq 1\ {\rm AU}$, reaches its maximal value at
the disk surface and then decreases.
This pronounced temperature rise at a disk surface can be explained by
the frequency-dependence of dust-absorption coefficient.
Absorption efficiency of dust particles is higher for short-wavelength
radiation than for long-wavelength radiation.
The disk surface, which is exposed directly to a high-temperature
radiation field in the halo, is therefore warmer than the disk interior
(Chiang, Goldreich 1997).

\begin{figure}
\begin{center}
\includegraphics[width=85mm,clip]{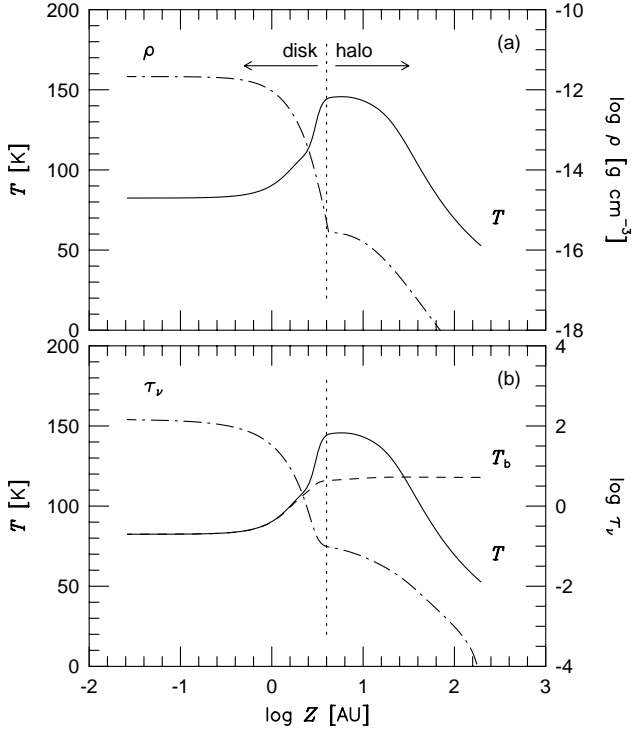}
\end{center}
\caption
{Vertical structure at $R = 10\ {\rm AU}$ for the standard model.
(a) Temperature $T$ (solid line) and density $\rho$ (dot-dashed line)
are plotted as a function of $Z$.
(b) Temperature $T$ (solid line), brightness temperature $T_{\rm b}$ at
$\nu = 10^{13}\ {\rm Hz}$ (dashed line), and optical depth $\tau_\nu$ at
$\nu = 10^{13}\ {\rm Hz}$ (dot-dashed line) measured from $Z = 200\ {\rm
AU}$ toward the midplane are plotted as a function of $Z$.}
\label{fig:fig2}
\end{figure}

In figure 2b the brightness temperature, $T_{\rm b}$, at frequency $\nu
= 10^{13}\ {\rm Hz}$ is shown as a function of $Z$.
We define $T_{\rm b}$ according to
$I_\nu = 2 h \nu^3 / c^2 [ \exp( h \nu / k_{\rm B} T_{\rm b} ) - 1 ]$,
where $I_\nu$ is the specific intensity, and is calculated assuming
that the system is viewed pole-on.
Also plotted in figure 2b is the optical depth, $\tau_\nu$, at $\nu =
10^{13}\ {\rm Hz}$, which is measured from $Z = 200\ {\rm AU}$ toward
the midplane.
As illustrated by the dashed line in figure 2b, $T_{\rm b}$ is equal to
$T$ near the midplane where $\tau_\nu \gg 1$; as $Z$ increases,
$T_{\rm b}$ also increases along with $T$ as long as $\tau_\nu \ge 1$.
When $\tau_\nu$ becomes less than unity, $T_{\rm b}$ begins to depart
from $T$, and keeps an almost constant value in the halo.
Thus, as expected, we find that the observed value of $T_{\rm b}$ is
determined by the value of $T$ at $\tau_\nu \simeq 1$.
Since the halo does not have $\tau_\nu$ larger than unity, we can define
the ``photosphere'' of the disk as the position of $\tau_\nu = 1$ at
mid- to far-infrared wavelengths.
Then, the observed brightness temperature at mid- to far-infrared
wavelengths traces the photospheric temperature of the disk.

\begin{figure}
\begin{center}
\includegraphics[width=85mm,clip]{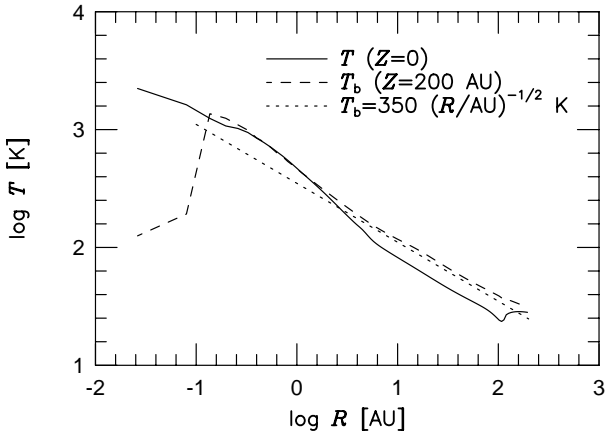}
\end{center}
\caption
{Radial distributions of the midplane temperature $T$ (solid line) and
the brightness temperature $T_{\rm b}$ at $\nu = 10^{13}\ {\rm Hz}$
(dashed line) evaluated at $Z = 200\ {\rm AU}$.
The dotted line represents a power law
$T_{\rm b}( R ) = 350\ (R/{\rm AU})^{-1/2}\ {\rm K}$.}
\label{fig:fig3}
\end{figure}

Figure 3 shows radial distributions of the midplane temperature and the
brightness temperature at $\nu = 10^{13}\ {\rm Hz}$ evaluated at $Z =
200\ {\rm AU}$.
We should be reminded that the SED is not determined by the midplane
temperature, but by the brightness temperature.
We find that it can be well approximated by
$T_{\rm b}( R ) = 350\ (R/{\rm AU})^{-1/2}\ {\rm K}$
in
$1\ {\rm AU} \le R \le 100\ {\rm AU}$.
This brightness temperature distribution is comparable to the
temperature distribution
$T( R ) = 307\ (R/{\rm AU})^{-0.49}\ {\rm K}$,
which was derived for HL~Tau through spectral modeling (Beckwith et al.\
1990).
This comparison shows that the disk--halo model can produce the large
mid- to far-infrared excess of the typical flat-spectrum T~Tauri star.
Moreover, from the result that the brightness temperature traces the
photospheric temperature of the disk, it follows that the mid- to
far-infrared excess originates from the disk.

\begin{figure}
\begin{center}
\includegraphics[width=85mm,clip]{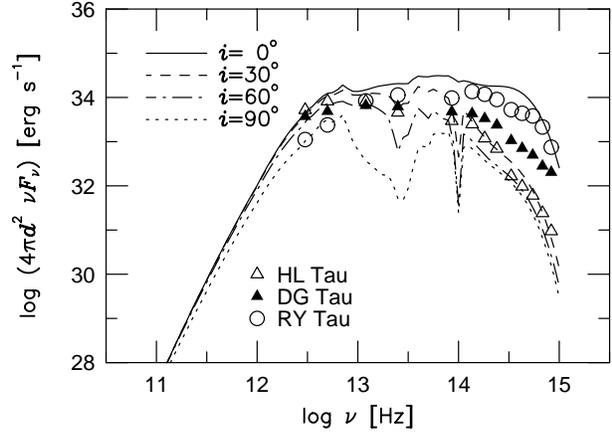}
\end{center}
\caption
{Spectral energy distributions of the standard model at four viewing
angles $i = 0^\circ$ (solid line), $30^\circ$ (dashed line), $60^\circ$
(dot-dashed line), and $90^\circ$ (dotted line).
The symbols represent observational data taken from Strom et al.\ (1989)
for HL~Tau (open triangles), DG~Tau (solid triangles) and RY~Tau (open
circles).}
\label{fig:fig4}
\end{figure}

In figure 4 the model SEDs are shown at several viewing angles.
As expected from the results described above, the model SEDs are nearly
flat in the infrared at viewing angles of $i = 0^\circ$ and $30^\circ$.
At $i = 60^\circ$, the model SED has a weak negative slope in the
infrared, with a deep absorption feature near $\lambda = 10\ \mu{\rm m}$
made by silicate and water ice (see Miyake, Nakagawa 1993).
The water ice feature at $3.1\ \mu{\rm m}$ is also seen at $i =
30^\circ$ and $60^\circ$.
In the optical and near-infrared, the shape of the model SED varies
significantly with the viewing angle.
The optical depth to the central star at $V$ band ($\lambda = 0.55\
\mu{\rm m}$) is 3, 18, and 45 at $i = 10^\circ$, $30^\circ$, and
$60^\circ$, respectively.
Thus, the central star is highly obscured when viewed at $i = 30^\circ$
and $60^\circ$, and the optical flux is due to the scattered light.

Observational data taken from Strom et al.\ (1989) for three
flat-spectrum T~Tauri stars (HL~Tau, DG~Tau, and RY~Tau) are also
plotted in figure 4, assuming that the distance to these sources is $d =
140\ {\rm pc}$.
Compared with the observational data, we find that the model SEDs can
reproduce the overall spectral shape of these sources, although the
model parameters are not tuned for any specific object.

\subsection{Effects of Disk Accretion}

\begin{figure}
\begin{center}
\includegraphics[width=85mm,clip]{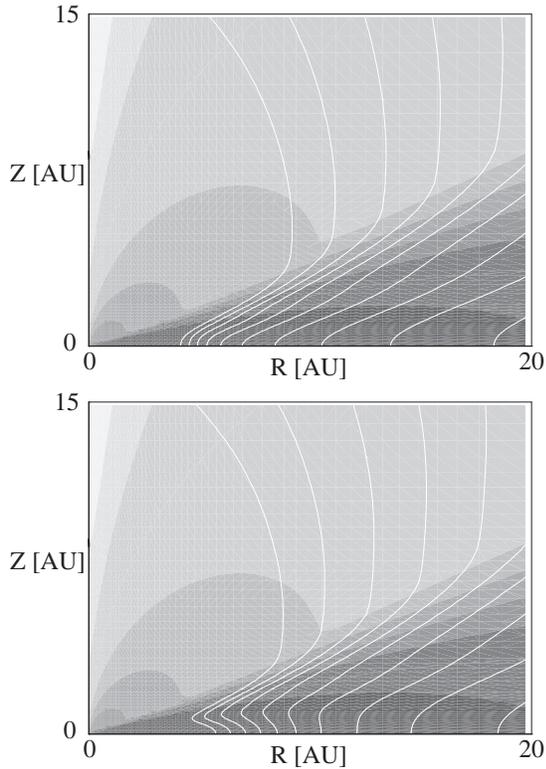}
\end{center}
\caption
{Closeup view of the density and temperature distributions to the
center for the standard model (upper panel) and for the steady accretion
model (lower panel).
For both models, the innermost contour is $T = 150\ {\rm K}$, and
contours decrease outward in $10\ {\rm K}$ intervals.}
\label{fig:fig5}
\end{figure}

In this subsection we examine the effects of disk accretion on the
temperature distribution and the SED.
The stellar luminosity is taken to be $L_\ast = 2\ \LO$, and the disk
accretion is assumed to contribute to a substantial fraction of the
total luminosity.
We first consider a steady accretion model, i.e., the accretion rate in
the disk is equal to that from the disk onto the star.
From analyses of the ultraviolet and optical excesses, the accretion
rate has been estimated to be in the range $5 \times 10^{-8} {\rm -} 5
\times 10^{-7}\ \MO\ {\rm yr}^{-1}$ for classical T~Tauri stars (Bertout
et al.\ 1988).
Hence, we adopt $\dot{M}_{\rm d} = \dot{M}_{\rm c} = 5 \times 10^{-7}\
\MO\ {\rm yr}^{-1}$ for the steady accretion model.
From equations ({\ref{eqn:Ld}) and ({\ref{eqn:Lc}), one obtains
accretion luminosities of $L_{\rm d} = 0.18\ \LO$ for the disk and
$L_{\rm c} = 2.46\ \LO$ for the bright spot, respectively.
The total luminosity is then $4.65\ \LO$, which is almost the same as
that of the standard model discussed in the previous subsection, and
thus it is possible to directly compare models with and without disk
accretion.

\begin{figure}
\begin{center}
\includegraphics[width=85mm,clip]{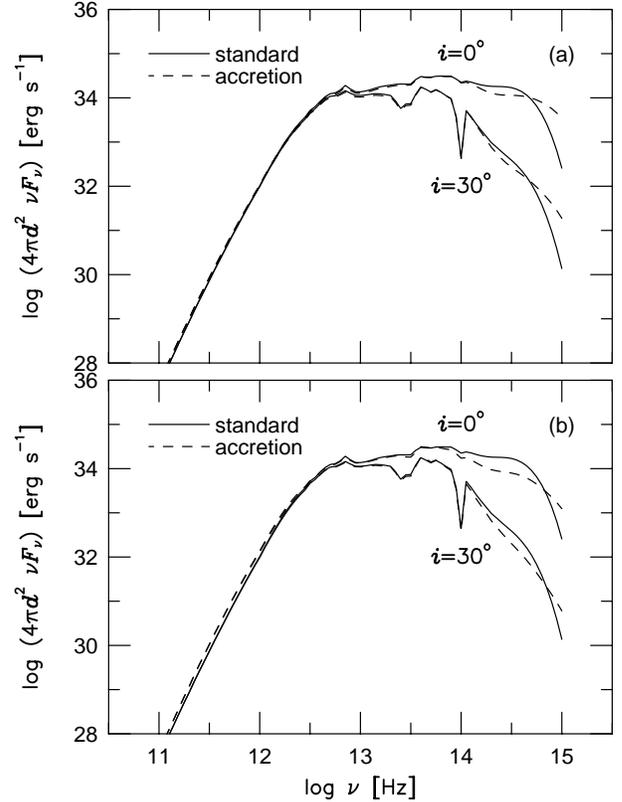}
\end{center}
\caption
{Effects of disk accretion on the SED.
The results are shown at two viewing angles, $i=0^\circ$ and $30^\circ$.
(a) Comparison of the steady accretion model (dashed line) with the
standard model (solid line).
(b) Comparison of the unsteady accretion model (dashed line) with the
standard model (solid line).}
\label{fig:fig6}
\end{figure}

Figure 5 shows a closeup view of the density and temperature
distributions to the center for the standard model (upper panel) and
for the steady accretion model (lower panel).
In the standard model, the temperature distribution is determined
entirely by radiative transport of energy emitted by the central star,
and the temperature monotonically increases along the vertical direction
in the disk.
On the other hand, the steady accretion model shows a more complicated
temperature structure in the disk.
Inside $\sim 10\ {\rm AU}$, the temperature is higher at the midplane in
the steady accretion model than in the standard model, indicating that
viscous dissipation effectively heats the disk in this region.
Although the temperature decreases along the vertical direction to
transport radiatively the energy generated by viscous dissipation, it
rises again toward the disk surface as in the standard model.
Note that the temperature distribution of the steady accretion model is
quite similar to that of the standard model in the halo and the disk
outside $\sim 10\ {\rm AU}$, suggesting that viscous dissipation is not
an important heating source in these regions.

Figure 6 compares the SEDs of models with and without disk accretion.
As discussed in the previous subsection, the mid- to far-infrared
portion of the SED is determined mainly by the photospheric temperature
of the outer disk.
It can be seen from figure 6a that the SED of the steady accretion model
does not differ significantly from that of the standard model in the
mid- to far-infrared region.
Therefore, the photospheric temperature of the outer disk is determined
by the radiation energy from the central star and the bright spot, not
by viscous dissipation in the disk.
The radiation energy source of the standard model is a 4000~K blackbody
with $5\ \LO$, while that of the steady accretion model is a combination 
of two blackbodies with different temperatures.
The difference in the radiation energy spectra appears in the
ultraviolet and optical portions of the SED.
Compared with the standard model, the steady accretion model shows more
ultraviolet and less optical fluxes.
The result shown in figure 6a indicates that the mid- to far-infrared
portion of the SED, and hence the photospheric temperature of the disk, 
is not sensitive to the spectrum of the central radiation energy source.
This is because the inner region of the halo immediately scatters or
absorbs/reradiates radiation from the center at longer wavelengths.

We next consider a model in which accretion takes place unsteadily, in
the sense that the disk accretion rate is constant in radius, but it is
not equal to the rate at which disk material flows from the inner edge
onto the central star.
The possibility that accretion flow in HL~Tau is unsteady was discussed
by Lin et al.\ (1994).
The flattened envelope around HL~Tau is inferred to accrete dynamically
onto the disk at a rate of $5 \times 10^{-6}\ \MO\ {\rm yr}^{-1}$, while
the accretion rate from the disk onto the central star should be smaller
than $7 \times 10^{-7}\ \MO\ {\rm yr}^{-1}$, because the accretion
luminosity cannot be larger than the total luminosity of HL~Tau (Lin et
al.\ 1994).
In the unsteady accretion model considered here, we adopt $\dot{M}_{\rm
d} = 5 \times 10^{-6}\ \MO\ {\rm yr}^{-1}$, which gives an intrinsic
disk luminosity of $L_{\rm d} = 1.81\ \LO$.
The accretion rate from the disk onto the central star is taken to be
$\dot{M}_{\rm c} = 2 \times 10^{-7}\ \MO\ {\rm yr}^{-1}$, which is one
order magnitude smaller than $\dot{M}_{\rm d}$, and produces a
bright-spot luminosity of $L_{\rm c} = 0.99\ \LO$.
The total luminosity then becomes $4.80\ \LO$, which is also comparable
to that of the standard model.

Figure 6b shows SEDs for the standard and unsteady accretion models.
Comparing these two models, we find no significant difference in the
mid- to far-infrared portion of the SED.
The unsteady accretion model, however, emits more ultraviolet and less
optical flux, which may be attributed to the difference in the spectra
of the central radiation energy sources.
This feature is similar to that seen in the steady accretion model.
In the unsteady accretion model, however, the central star and bright
spot emit only about $3\ \LO$ to heat the disk, which would not be
sufficient to produce a mid- to far-infrared flux as large as that of
the standard or steady accretion model.
This deficit of radiation energy is compensated by the accretion energy
generated in the innermost region of the disk.
In general, most of the accretion energy is released in the innermost
region of the system where the gravitational potential well is deep.
In the unsteady accretion model we adopt, equation (\ref{eqn:D})
indicates that 80\% of the intrinsic disk luminosity is generated inside
$1.2\ {\rm AU}$.
The accretion energy released in the innermost region of the disk is
radiated at short wavelengths.
This emission is subsequently scattered or absorbed and reradiated by
the halo to heat the outer region of the disk, producing the large mid-
to far-infrared excess, as shown in figure 6b.

Figure 6b also shows that the unsteady accretion model produces more
flux at radio wavelengths than the standard model.
At radio wavelengths, our disk model is optically thin, and thus the
emergent flux is related to the temperature at the midplane where disk
material is concentrated.
Our result indicates that, for a fixed total luminosity, disk accretion
at a rate as high as $5 \times 10^{-6}\ \MO\ {\rm yr}^{-1}$ can increase
the radio flux by increasing the midplane temperature, although it does
not modify the mid- to far-infrared portion of the SED.

\begin{figure}
\begin{center}
\includegraphics[width=85mm,clip]{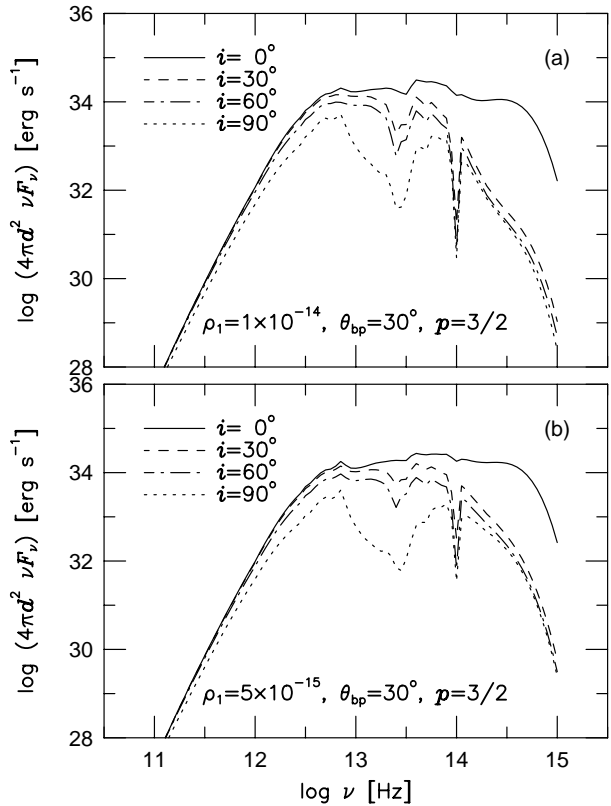}
\end{center}
\caption
{Effects of the size of the bipolar lobes on the SED.
(a) SEDs for a model with
$\rho_1 = 1.0 \times 10^{-14}\ {\rm g~cm}^{-3}$,
$\theta_{\rm bp} = 30^\circ$,
and
$p = 3/2$.
(b) SEDs for a model with
$\rho_1 = 5.0 \times 10^{-15}\ {\rm g~cm}^{-3}$,
$\theta_{\rm bp} = 30^\circ$,
and
$p = 3/2$.}
\label{fig:fig7}
\end{figure}

\subsection{Dependence on the Halo Density Distribution}

\begin{figure}
\begin{center}
\includegraphics[width=85mm,clip]{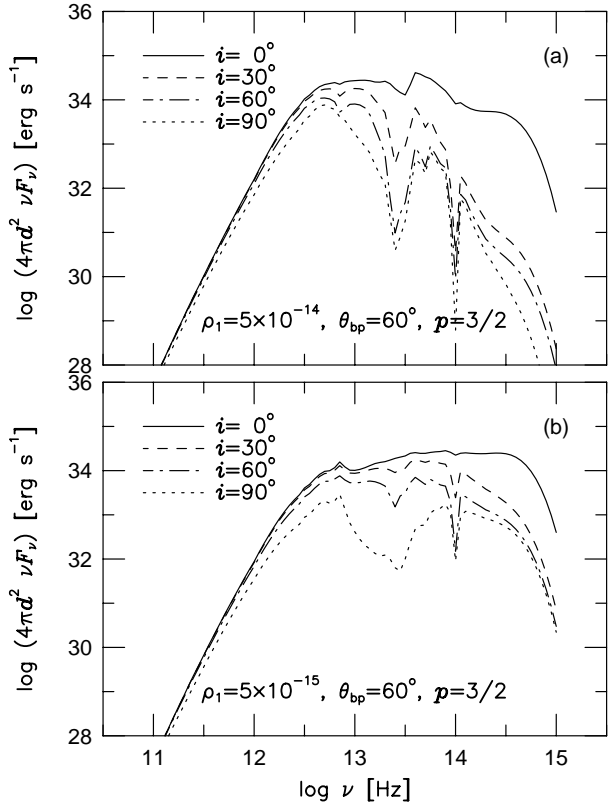}
\end{center}
\caption
{Dependence of the SED on the reference density $\rho_1$.
(a) SEDs for a model with
$\rho_1 = 5.0 \times 10^{-14}\ {\rm g~cm}^{-3}$,
$\theta_{\rm bp} = 60^\circ$,
and
$p = 3/2$.
(b) SEDs for a model with
$\rho_1 = 5.0 \times 10^{-15}\ {\rm g~cm}^{-3}$,
$\theta_{\rm bp} = 60^\circ$,
and
$p = 3/2$.}
\label{fig:fig8}
\end{figure}

\begin{figure}
\begin{center}
\includegraphics[width=85mm,clip]{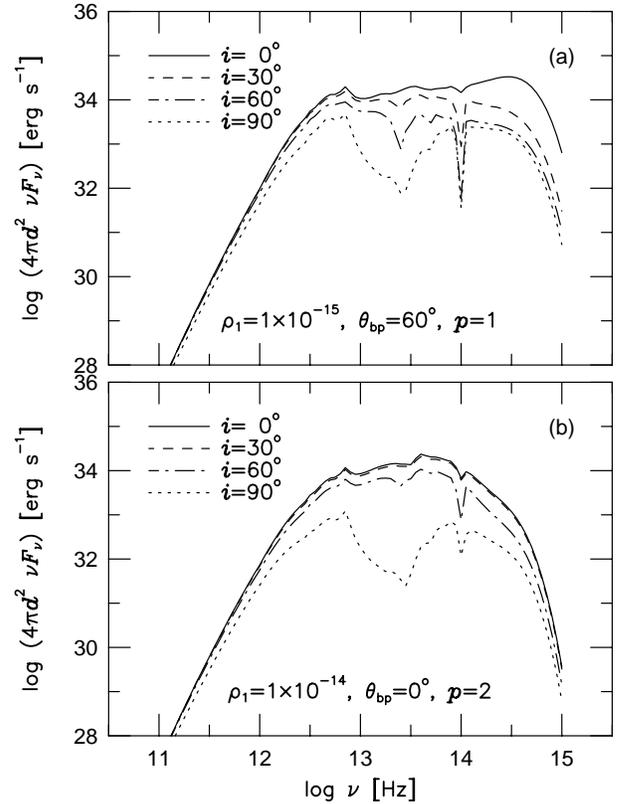}
\end{center}
\caption
{Dependence of the SED on the power-law index, $p$.
(a) SEDs for a model with
$p = 1$,
$\rho_1 = 1.0 \times 10^{-15}\ {\rm g~cm}^{-3}$,
and
$\theta_{\rm bp} = 60^\circ$.
(b) SEDs for a model with
$p = 2$,
$\rho_1 = 1.0 \times 10^{-14}\ {\rm g~cm}^{-3}$,
and
$\theta_{\rm bp} = 0^\circ$.}
\label{fig:fig9}
\end{figure}

As illustrated in the previous subsection, the mid- to far-infrared
portion of the SED is not sensitive to disk accretion.
We therefore examine how the flat infrared spectrum depends on the
assumed halo density distribution using models without disk accretion.
We first test the sensitivity of the SED to the size of the bipolar
lobes.
Figure 7a shows SEDs for a model with the opening half-angle of the
bipolar holes being $\theta_{\rm bp} = 30^\circ$.
The other parameters for the halo density distribution are the same as
those of the standard model, i.e., $\rho_1 = 10^{-14}\ {\rm g~cm}^{-3}$
and $p = 3/2$.
Compared with the standard model having $\theta_{\rm bp} = 60^\circ$,
the emergent flux is considerably reduced at optical and near-infrared
wavelengths when viewed at $i \ge 30^\circ$, because less optical and
near-infrared light can escape from the central star through the
bipolar holes.
This means that more energy from the central star is used to heat the
outer region of the disk.
As a result, the mid- to far-infrared flux increases slightly, compared
with the standard model.
However, the effect of narrowing the opening half-angle of the bipolar
holes may be compensated by reducing $\rho_1$.
Figure 7b shows SEDs for a model with a lower halo density: $\rho_1 = 5
\times 10^{-15}\ {\rm g~cm}^{-3}$, $\theta_{\rm bp} = 30^\circ$, and $p
= 3/2$.
As with the standard model, this model also produces flat infrared
spectra when the viewing angle is $0^\circ \le i \le 60^\circ$.
However, some differences are found in the SED.
The absorption feature near $10\ \mu{\rm m}$ is shallower due to the
reduced halo density, and the slope of the SED is more steep from the
near-infrared to the optical.

Secondly, we examine how the halo density affects the SED.
In figures 8a and 8b, model SEDs are shown for $\rho_1 = 5 \times
10^{-14}$ and $5 \times 10^{-15}\ {\rm g~cm}^{-3}$, respectively.
For both models, we adopt $\theta_{\rm bp} = 60^\circ$ and $p=3/2$.
The model with a higher halo density, $\rho_1 = 5 \times 10^{-14}\
{\rm g~cm}^{-3}$, shows relatively narrow SEDs when viewed at $i > 0$,
and the peak of the SED occurs in the far-infrared.
The emergent flux steeply declines from the near-infrared to the
optical, because light from the hot inner regions is highly
extinguished by the dense halo and only a small amount of scattered
photon can escape the halo.
Thus, this model is not classified as a ``class II'' object with
flat infrared spectrum unless it is observed precisely pole-on.
On the other hand, a model with a lower halo density,
$\rho_1 = 5 \times 10^{-15}\ {\rm g~cm}^{-3}$, still produces flat SEDs
at a wide range of viewing angles, $\sim 30^\circ {\rm -} 60^\circ$.

Finally, we examine the dependence of SEDs on the power-law index, $p$.
A molecular cloud core which collapses to form a star and a disk is
expected to be rotating.
At later stages of the collapse, infalling material lands onto the
equatorial plane, rather than onto the central star.
Thus, the halo density is reduced inside the centrifugal radius, and it
decreases more slowly with the radius than spherically symmetric cases.
We then consider a model with $p = 1$, $\rho_1 = 10^{-15}\ {\rm
g~cm}^{-3}$, and $\theta_{\rm bp} = 60^\circ$.
As shown in figure 9a, this model also produces flat infrared spectra
at viewing angles of $i \sim 30^\circ {\rm -} 60^\circ$.
Compared with the standard model, it emits more optical flux, because
the halo density is lower in the innermost region, so that more optical
light escapes the halo.

If the central star drives a stellar wind with constant velocity and
mass outflow rate, a power law, $\rho \propto r^{-2}$, is appropriate
for the halo density distribution.
Figure 9b shows SEDs for a model with $p = 2$, $\rho_1 = 10^{-14}\ {\rm
g~cm}^{-3}$, and $\theta_{\rm bp} = 0^\circ$.
Because the slope of the SEDs is positive in the infrared, $p = 2$ is
not favored by the flat spectrum.

\section{Discussion}

\subsection{Origin of the Flat Spectrum}

The present results show that the disk--halo model can reproduce the
overall spectral shape of typical flat-spectrum T~Tauri stars.
In our model, the mid- to far-infrared excesses of flat-spectrum T~Tauri
stars originate from the disk, rather than from the halo.
The halo needed to heat the disk can be as compact as the disk, itself,
and we suggest that a reflection nebula often associated with a T~Tauri
star is an observational counterpart of the halo.
The reflection nebula probably corresponds to the inner part of a
remnant of an infalling envelope.
The disk--halo model is therefore consistent with the notion that the
flat-spectrum T~Tauri stars are in a transitional stage from protostars
to T~Tauri stars (Hayashi et al.\ 1993).

In contrast with the disk--halo model, the infalling envelope model of
Calvet et al.\ (1994) explains the mid- to far-infrared excesses by the
emission from infalling envelopes.
Indeed, observations of molecular line emission have revealed extended,
disk-like envelopes associated with several flat-spectrum T~Tauri stars
(Hayashi et al.\ 1993; Kitamura et al.\ 1996a; Momose et al.\ 1996).
However, recent high-resolution observations have shown that the
continuum emission at submillimeter and millimeter wavelengths
originates entirely from a compact ($\ltsim 100\ {\rm AU}$) region,
suggesting that the extended ($\sim 1000\ {\rm AU}$) envelopes traced
by molecular line emission do not contribute to the submillimeter and
millimeter emission (Lay et al.\ 1994; Kitamura et al.\ 1996b; Looney et
al.\ 2000).
Since infrared emission originates from warmer regions than
submillimeter and millimeter emission, it is quite unlikely that the
extended envelopes contribute to the infrared emission.
Hence, the mid- to far-infrared excesses of flat-spectrum T~Tauri stars
should also originate from a compact region.
The disk--halo model produces emission at any wavelength from a compact
region, and therefore is more consistent with this observational
constraint than the infalling envelope model.

If the mid- to far-infrared excesses of flat-spectrum T~Tauri stars are
to be explained by the emission arising at $100 {\rm -} 1000\ {\rm AU}$
in their infalling envelopes (Calvet et al.\ 1994), they should have
been surrounded by sufficient amount of infalling material at a radius
of $100 {\rm -} 1000\ {\rm AU}$ as protostars are.
Growing observational evidence, however, indicates that at these radii
the amount of infalling material around flat-spectrum T~Tauri stars
is much less than that around embedded protostars (Ohashi et al.\ 1991,
1996), suggesting that the amount of infalling material
inside $100 {\rm -} 1000\ {\rm AU}$ around the central sources decreases
during the course of evolution from protostars to T~Tauri stars.
Dispersing motion in the envelopes has actually been observed for T~Tau
(Momose et al.\ 1996) and DG~Tau (Kitamura et al.\ 1996a).
HL~Tau is suggested to be situated in the bubble wall, which is an
expanding shell with XZ~Tau being the closest known source to the center
(Welch et al.\ 2000).
In addition, Momose (1998) showed that envelopes around flat-spectrum
T~Tauri stars are less centrally condensed than those around protostars.
These results imply that infalling material around $100 {\rm -} 1000\
{\rm AU}$ has been dissipated away, or has accreted onto the central
star/disk system.

In our view, the inner parts of the envelopes around flat-spectrum
T~Tauri stars are still optically thick to stellar radiation, although
they do not have sufficient mass to produce the mid- to far-infrared
excesses.
For instance, the visual extinction toward HL~Tau was estimated to be
$A_V > 22\ {\rm mag}$ (Stapelfeldt et al.\ 1995), which is comparable to
that in our calculations.
In the disk--halo model, visual extinction varies widely depending on
the viewing angle as well as on the density distribution in the halo;
thus, our model can explain the flat spectrum for objects with various
visual extinction values.
The detailed structure of the halo, however, cannot be predicted by the
analysis of SEDs, because the halo structure is not unique to give the
observed flat spectra.
Obviously, to solve the degeneracy, further observational information is
needed on the structure within 100 AU.
High-resolution observations of the scattered light in the near-infrared
may provide crucial information on the structure of the disk and halo
(Close et al.\ 1997).
We will report on a comparison between the disk--halo model and
near-infrared observations in a future paper.

\subsection{Effects of Shock Heating}

If the halo is an innermost part of the infalling envelope, the envelope
gas should shock the disk surface.
This shock heating could be an important energy source which our analyses
do not take into account.
To clarify quantitatively how shock heating modifies the SED in the
infrared region, detailed radiation hydrodynamics calculations will be
needed.
Although such analysis is beyond the scope of this paper, we can discuss
that the amount of energy released at the disk surface by the shock
would not be sufficient to modify the SED obtained in our standard
model.

Suppose that envelope gas freely falls onto the disk along a ballistic
orbit, the initial cloud core is spherically symmetric and uniformly
rotating, the infall is initiated by the expansion wave (Shu 1977), and
the gas joins a Keplerian disk behind the shock.
The amount of energy dissipated at shock per unit mass is
$GM_\ast/2R$.
We denote by $F(R)$ the mass flux from the envelope on the disk at
radius $R$, which is given by equation (2.1) in Nakamoto and Nakagawa
(1994).
If we adopt $\dot{M} = 0.9 \times 10^{-5} \MO\ {\rm yr}^{-1}$ (Hayashi
et al.\ 1993), $R_{\rm out} = 100\ {\rm AU}$ ($R_{\rm out}$ is
denoted by $R_d$ in Nakamoto, Nakagawa 1994), and $M_\ast = 0.5 \MO$,
the energy generation rate due to shock heating at the disk surface per
unit area, $\varepsilon_{\rm s}$, is given by
\begin{eqnarray}
\varepsilon_{\rm s} & = & \frac{1}{2} \frac{GM_\ast}{R} F(R) \nonumber \\
& = & \frac{1}{2} \frac{GM_\ast}{R_{\rm out}}
\left( \frac{R_{\rm out}}{R} \right)^2
{ \frac{1}{\sqrt{ 1 - R/R_{\rm out}} } } \nonumber \\
& = & 0.45 \left( \frac{R}{\rm 100~AU} \right)^{-2}
\left( 1 - R/{\rm 100~AU} \right)^{-1/2} \nonumber \\
&   & \hspace{4cm} {\rm erg~cm}^{-2}\ {\rm s}^{-1}~.
\end{eqnarray}
If there were no irradiation from the central star, the disk surface
would be heated by shock heating to the temperature
\begin{eqnarray}
T_{\rm eff} & = & \left( \varepsilon_{\rm s} / \sigma_{\rm B} \right)^{1/4}
\nonumber \\
& = & 94 \left( \frac{R}{\rm 1~AU} \right)^{-1/2}
\left( 1-R/{\rm 100~AU} \right)^{-1/8}~{\rm K}~,
\end{eqnarray}
which is much lower than the temperature obtained by our standard model
including irradiation (see figure 3).
Therefore, it seems that shock heating at the surface of the disk has
only a negligible effect on the observed SED from the flat-spectrum
T~Tauri disk.

\section{Conclusions}

\begin{enumerate}
\item We have shown that disks heated by the scattering and reprocessing of
the stellar radiation through the halo can have flat infrared spectra.

\item Local viscous heating is not sufficient to produce large mid- to
far-infrared emission from the disk if we consider a reasonable rate of
mass accretion in disks around classical T~Tauri stars.
However, the accretion energy released in the innermost region of the
star/disk system is transported by radiation through the halo to heat
the outer region of the disk, resulting again in a flat spectrum.

\item We examined the sensitivity of the SED to the assumed halo
structure, and found that density distributions with a power-law index
$\le 3/2$ can provide the backwarming needed for flat infrared spectra.
However, we have found that it is difficult to constrain the halo
structure only from the SED in the infrared.

\item We have discussed that the halo will be observed as a reflection
nebula often associated with a T~Tauri star, indicating that it is the
inner part of a remnant of an infalling envelope.
A more detailed test of the model will be made by comparisons with
imaging observations of near-infrared scattered light.
\end{enumerate}

\vspace{1pc}

We are grateful to M.\ Hayashi and M.\ Umemura for valuable discussions.
We also thank an anonymous referee for suggestions which improved the
paper.
The computations were performed on CP-PACS at the Center for
Computational Physics in University of Tsukuba, and on the Fujitsu
VPP300/16R at the Astronomical Data Analysis Center of the National
Astronomical Observatory, Japan.
TN was partially supported by the Grant-in-Aid for Scientific Research
on Priority Areas (10147105) and for Scientific Research (10740093) of
the Ministry of Education, Culture, Sports, Science, and Technology,
Japan.



\begin{thebibliography}{}
\bibitem[Adams, Lada, Shu 1988]{ALS1988}
    Adams, F.~C.,
    Lada, C.~J., \&
    Shu, F.~H.\
    1988, ApJ, 326, 865
\bibitem[Adams, Shu 1986]{AS1986}
    Adams, F.~C., \&
    Shu, F.~H.\
    1986, ApJ, 308, 836
\bibitem[Beckwith et al.\ 1990]{BSCG1991}
    Beckwith, S.~V.~W.,
    Sargent, A.~I.,
    Chini, R.~S., \&
    G\"usten, R.\
    1990, AJ, 99, 924
\bibitem[Bertout, Basri, Bouvier 1988]{BBB1988}
    Bertout, C.,
    Basri, G., \&
    Bouvier, J.\
    1988, ApJ, 330, 350
\bibitem[Butner, Natta, Evans 1994]{BNE1994}
    Butner, H.~M.,
    Natta, A., \&
    Evans, N.~J., II
    1994, ApJ, 420, 326
\bibitem[Calvet et al.\ 1994]{CHKW1994}
    Calvet, N.,
    Hartmann, L.,
    Kenyon, S.~J., \&
    Whitney, B.~A.\
    1994, ApJ, 434, 330
\bibitem[Chiang, Goldreich 1997]{CG1997}
    Chiang, E.~I., \&
    Goldreich, P.\
    1997, ApJ, 490, 368
\bibitem[Close et al.\ 1997]{CRNRG1997}
    Close, L.~M.,
    Roddier, R.,
    Northcott M.~J.,
    Roddier, C., \&
    Graves, J.~E.\
    1997, ApJ, 478, 766
\bibitem[D'Alessio, Calvet, Hartmann 1997]{DCH1997}
    D'Alessio, P.,
    Calvet, N., \&
    Hartmann, L.\
    1997, ApJ, 474, 397
\bibitem[Hayashi, Nakagawa, Nakazawa 1985]{HNN1985}
    Hayashi, C.,
    Nakazawa, K., \&
    Nakagawa, Y.\
    1985,
    in Protostars and Planets II,
    ed.\ D.~C.\ Black \& M.~S.\ Matthews
    (Tucson: Univ.\ Arizona Press), 1100
\bibitem[Hayashi, Ohashi, Miyama 1993]{HOM1993}
    Hayashi, M.,
    Ohashi, N., \&
    Miyama, S.~M.\
    1993, ApJ, 418, L71
\bibitem[Kenyon, Calvet, Hartmann 1993]{KCH1993}
    Kenyon, S.~J.,
    Calvet, N., \&
    Hartmann, L.\
    1993, ApJ, 414, 676
\bibitem[Kenyon, Hartmann 1987]{KH1987}
    Kenyon, S.~J., \&
    Hartmann, L.\
    1987, ApJ, 323, 714
\bibitem[Kitamura, Kawabe, Saito 1996a]{KKS1996a}
    Kitamura, Y.,
    Kawabe, R., \&
    Saito, M.\
    1996a, ApJ, 457, 277
\bibitem[Kitamura, Kawabe, Saito 1996b]{KKS1996b}
    Kitamura, Y.,
    Kawabe, R., \&
    Saito, M.\
    1996b, ApJ, 465, L137
\bibitem[Kusaka, Nakano, Hayashi 1970]{KNH1970}
    Kusaka, T.,
    Nakano, T., \&
    Hayashi, C.\
    1970, Prog.\ Theor.\ Phys.\ 44, 1580
\bibitem[Lay et al.\ 1994]{LCHP1994}
    Lay, O.~P.,
    Carlstrom, J.~E.,
    Hills, R.~E., \&
    Phillips, T.~G.\
    1994, ApJ, 434, L75
\bibitem[Lin et al.\ 1994]{LHBO1994}
    Lin, D.~N.~C.,
    Hayashi, M.,
    Bell, K.~R., \&
    Ohashi, N.\
    1994, ApJ, 435, 821
\bibitem[Looney, Mundy, Welch 2000]{LMW2000}
    Looney, L.~W.,
    Mundy, L.~G., \&
    Welch, W.~J.\
    2000, ApJ, 529, 477
\bibitem[Lynden-Bell, Pringle 1974]{LBP1974}
    Lynden-Bell, D., \&
    Pringle, J.~E.\
    1974, MNRAS, 168, 603
\bibitem[Masunaga, Miyama, Inutsuka 1998]{MMI1998}
    Masunaga, H.,
    Miyama, S.~M., \&
    Inutsuka, S.\
    1998, ApJ, 495, 346
\bibitem[Miyake, Nakagawa 1993]{MN1993}
    Miyake, K., \&
    Nakagawa, Y.\
    1993, Icarus, 106, 20
\bibitem[Momose 1998]{M1998}
    Momose, M.\
    1998, PhD thesis, The Graduate University for Advanced Studies
\bibitem[Momose et al.\ 1996]{MOKHN1996}
    Momose, M.,
    Ohashi, N.,
    Kawabe, R.,
    Hayashi, M., \&
    Nakano, T.\
    1996, ApJ, 470, 1001
\bibitem[Nakamoto, Nakagawa 1994]{NN1994}
    Nakamoto, T., \&
    Nakagawa, Y.\
    1994, ApJ, 421, 640
\bibitem[Natta 1993]{N1993}
    Natta, A.\
    1993, ApJ, 412, 761
\bibitem[Ohashi et al.\ 1996]{OHKI1996}
    Ohashi, N.,
    Hayashi, M.,
    Kawabe, R., \&
    Ishiguro, M.\
    1996, ApJ, 466, 317
\bibitem[Ohashi et al.\ 1991]{OKHI1991}
    Ohashi, N.,
    Kawabe, R.,
    Hayashi, M., \&
    Ishiguro, M.\
    1991, AJ, 102, 2054
\bibitem[Sargent, Beckwith 1991]{SB1991}
    Sargent, A.~I., \&
    Beckwith, S.~V.~W.\
    1991, ApJ, 382, L31
\bibitem[Shu 1977]{S1977}
    Shu, F.~H.\
    1977, ApJ, 214, 488
\bibitem[Stapelfeldt et al.\ 1995]{S1995}
    Stapelfeldt, K.~R.,
    Burrows, C.~J.,
    Krist, J.~E,,
    Trauger, J.~T.,
    Hester, J.~J.,
    Holtzman, J.~A.,
    Ballester, G.~E.,
    Casertano, S.,
    Clarke, J.~T.
    Crisp, D.,
    Evans, R.~W.,
    Gallagher, J.~S., III,
    Griffiths, R.~E.,
    Hoessel. J.~G.,
    Mould, J.~R.,
    Scowen, P.~A.,
    Watson, A.~M., \&
    Westphal, J.~A.\
    1995, ApJ, 449, 888
\bibitem[Stone, Mihalas, Norman 1992]{SMN1992}
    Stone, J.~M.,
    Mihalas, D., \&
    Norman, M.~L.\
    1992, ApJS, 80, 819
\bibitem[Strom et al.\ 1989]{SSECS1989}
    Strom, K.~M.,
    Strom, S.~E.,
    Edwards, S.,
    Cabrit, S., \&
    Skrutskie, M.~F.\
    1989, AJ, 97, 1451
\bibitem[Welch et al.\ 2000]{WHHB2000}
    Welch, W.~J.,
    Hartmann, L.,
    Helfer, T., \&
    Brice\~no, C.\
    2000, ApJ, 540, 362
\end{thebibliography}
\end{document}